\documentclass[%
 reprint,
superscriptaddress,
 amsmath,amssymb,
 aps,
]{revtex4-1}

\usepackage{graphicx}
\usepackage{dcolumn}
\usepackage{bm}
\usepackage{amsmath,amssymb,amstext,amsthm}
\usepackage{braket}
\usepackage{bbold}
\usepackage{bbm}
\usepackage{xcolor}
\usepackage{ gensymb }
\usepackage{soul}
\usepackage{enumitem}

\begin{document}

\preprint{APS/123-QED}

\title{Methods for preparation and detection of neutron spin-orbit states}

\author{D. Sarenac}
\email{dsarenac@uwaterloo.ca}
\affiliation{Department of Physics, University of Waterloo, Waterloo, ON, Canada, N2L3G1}
\affiliation{Institute for Quantum Computing, University of Waterloo,  Waterloo, ON, Canada, N2L3G1}
\author{J. Nsofini} 
\affiliation{Department of Physics, University of Waterloo, Waterloo, ON, Canada, N2L3G1}
\affiliation{Institute for Quantum Computing, University of Waterloo,  Waterloo, ON, Canada, N2L3G1}
\author{I. Hincks} 
\affiliation{Institute for Quantum Computing, University of Waterloo,  Waterloo, ON, Canada, N2L3G1}
\affiliation{Department of Applied Math, University of Waterloo, Waterloo, ON, Canada, N2L3G1}

\author{M. Arif}
\affiliation{National Institute of Standards and Technology, Gaithersburg, Maryland 20899, USA}
\author{Charles W. Clark}
\affiliation{Joint Quantum Institute, National Institute of Standards and Technology and University of Maryland, College Park, Maryland 20742, USA}
\author{D. G. Cory}
\affiliation{Institute for Quantum Computing, University of Waterloo,  Waterloo, ON, Canada, N2L3G1} 
\affiliation{Department of Chemistry, University of Waterloo, Waterloo, ON, Canada, N2L3G1}
\affiliation{Perimeter Institute for Theoretical Physics, Waterloo, ON, Canada, N2L2Y5}
\affiliation{Canadian Institute for Advanced Research, Toronto, Ontario, Canada, M5G 1Z8}
\author{M. G. Huber}
\affiliation{National Institute of Standards and Technology, Gaithersburg, Maryland 20899, USA}
\author{D. A. Pushin}
\email{dmitry.pushin@uwaterloo.ca}
\affiliation{Department of Physics, University of Waterloo, Waterloo, ON, Canada, N2L3G1}
\affiliation{Institute for Quantum Computing, University of Waterloo,  Waterloo, ON, Canada, N2L3G1}

\date{\today}

\begin{abstract}
The generation and control of neutron orbital angular momentum (OAM) states and spin correlated OAM (spin-orbit) states provides a powerful probe of materials with unique penetrating abilities and magnetic sensitivity. We describe techniques to prepare and characterize neutron spin-orbit states, and provide a quantitative comparison to known procedures. The proposed detection method directly measures the correlations of spin state and transverse momentum, and overcomes the major challenges associated with neutrons, which are low flux and small spatial coherence length. Our preparation techniques, utilizing special geometries of magnetic fields, are based on coherent averaging and spatial control methods borrowed from nuclear magnetic resonance. The described procedures may be extended to other probes such as electrons and electromagnetic waves. 
\end{abstract}

\pacs{Valid PACS appear here}
\maketitle

\section{\label{sec:level1}Introduction}

In addition to possessing spin angular momentum, beams of light \cite{LesAllen1992}, electrons \cite{uchida2010generation,mcmorran2011electron}, and neutrons \cite{Dima2015,holography} can carry orbital angular momentum (OAM) parallel to their propagation axis. There have been many recent developments in preparation and detection of OAM waves \cite{BarnettBabikerPadgett,rubinsztein2016roadmap}, and they have found numerous applications in microscopy, encoding and multiplexing of communications, quantum information processing, and the manipulation of matter \cite{mair2001entanglement,wang2012terabit,Andersen2006,he1995direct,friese1996optical,brullot2016resolving}.

In addition, it is possible to create ``spin-orbit'' states in which the spin and orbital angular momentum are correlated. For light, the correlation is between OAM and the polarization degree of freedom (DOF) \cite{maurer2007tailoring,qplate}, while for electrons and neutrons it is between OAM and the spin DOF \cite{karimi2012spin,spinorbit}. Optical spin-orbit beams have demonstrated a number of applications in high resolution optical imaging, high-bandwidth communication, optical metrology, and quantum cryptography \cite{marrucci2011spin,milione20154,schmiegelow2016transfer,vallone2014freespace}.

In this paper we develop methods of producing neutron spin-orbit states using special geometries of magnetic fields. The techniques are based on coherent averaging and spatial control methods borrowed from nuclear magnetic resonance \cite{zhang1995analysis,cory1989chemical,sodickson1998generalized,levitt_composite_1986}. We then quantify and compare the practical methods for preparation and detection of neutron spin-orbit states. Lastly, we propose a method to characterize neutron spin-orbit states by measuring correlations between the spin direction and the momentum projected to a specific axis. This detection technique may be used to overcome the main challenges associated with low flux and the small spatial coherence of neutron beams.

\section{OAM Preparation with a Spiral Phase Plate}

\begin{figure*}
\center
\includegraphics[width=\linewidth]{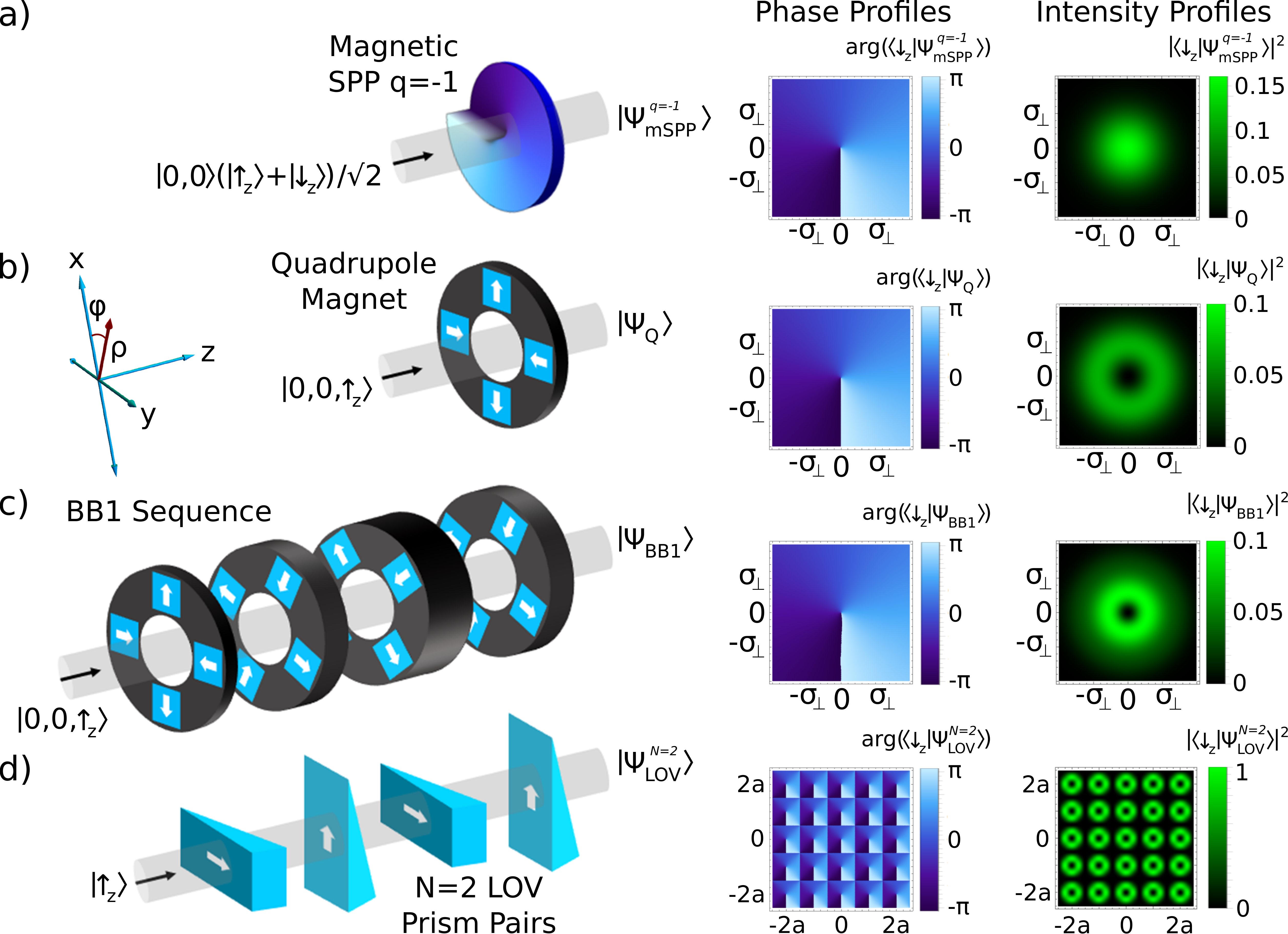}
\caption{Four methods of producing neutron spin-orbit states. The phase and intensity profiles of the output states, post-selected on the spin state correlated to the OAM, are shown on the right. a) An incoming neutron wavepacket in a coherent superposition of the two spin eigenstates passes through a magnetic SPP which is made out of a material with equal magnetic and nuclear scattering lengths, thereby inducing an azimuthally varying phase for only one spin state. b) A spin-polarized neutron wavepacket passes through a quadrupole magnetic field which induces the spin-orbit state \cite{spinorbit}. After transversing the quadrupole field, the intensity profile of the spin state correlated to the OAM has a ring shape. c) A sequence of quadrupoles with appropriate length and orientation acts as a BB1 pulse which increases the radii at which the spin and OAM are maximally entangled. d) In analogy to the LOV prism pairs capable of generating lattices of optical spin-orbit states \cite{sarenac2017generation}, a sequence of magnetic prisms can be used to approximate the quadrupole operator and produce a lattice of neutron spin-orbit states.}
 \label{Fig:methods}
\end{figure*}

A direct way of generating OAM waves is to pass a Gaussian beam through an azimuthally varying potential gradient such as that of a spiral phase plate (SPP) \cite{beijersbergen1994helical}. Here we examine a scenario in which a coherent neutron wavepacket is traveling on axis with the SPP. It is convenient to consider a neutron traveling along the $\hat{z}$ direction with momentum $\hbar k_z$ and with equal transverse spatial coherence lengths ($\sigma_x=\sigma_y\equiv \sigma_\bot$, where $\sigma_{x,y}=1/(2\Delta k_{x,y})$ and $\Delta k_{x,y}$ are the spreads of the wavepacket's transverse wavevectors). The transverse eigenstates can then be conveniently expressed in cylindrical coordinates $(\rho,\phi,z)$ as:

\begin{align}
\ket{n,\ell,s}=\mathcal{N} \xi^{|\ell|}e^{-\frac{\xi^2}{2}}\mathcal{L}_{n}^{|\ell|}\left(\xi^2\right)e^{i\ell\phi}\ket{s},
\label{basis}
\end{align}

\noindent where 
$\mathcal{N}$
is a normalization constant, $\xi=\rho/\sigma_\perp$ is the rescaled radial coordinate, 
$n\in\{0,1,2...\}$ is the radial quantum number, $\ell\in \{0, \pm 1, \pm2...\} $ is the azimuthal quantum number indicative of OAM, $\mathcal{L}_{n}^{|\ell|}\left(\xi^2\right)$ are the associated Laguerre polynomials, and $s\in\{\uparrow,\downarrow\}$ describes the spin state. Applying the OAM operator $\hat{L}_z=-i\hbar\frac{\partial}{\partial\phi}$ to Eq.~\ref{basis} verifies that this wavepacket carries an OAM of $\ell \hbar$ parallel to its propagation axis. 

An SPP provides an azimuthal potential gradient which induces OAM relative to the SPP axis. The thickness profile of an SPP is given by $h(\phi)=h_\mathrm{0}+h_\mathrm{s}\phi/(2\pi)$, where $h_\mathrm{0}$ is the base thickness and $h_\mathrm{s}$ is the thickness of the step. In neutron optics \cite{Sears}, a wavepacket propagating on axis through an SPP acquires a spatially dependent phase  $\alpha(\phi)=-Nb_\mathrm{c}\lambda h(\phi)=\alpha_\mathrm{0}+q\phi$, where $N b_\mathrm{c}$ is the coherent scattering length density of the SPP material, $\lambda$ is the neutron de Broglie wavelength, $q=-Nb_\mathrm{c}\lambda h_\mathrm{s}/(2\pi)$  is known as the topological charge or the winding number of the SPP \cite{Topphase}, and $\alpha_\mathrm{0}=-Nb_\mathrm{c}\lambda h_\mathrm{0}$ is the phase shift associated with the base thickness. The effect of the SPP on the neutron wavefront can be expressed as an operator:

\begin{align}
\hat{U}_{\mathrm{SPP}}=e^{i\alpha_\mathrm{0}}e^{iq\phi}.
\end{align}

For example, consider an incoming neutron wavepacket with a definite value of OAM:

\begin{align}
\ket{\Psi_\mathrm{in}}=\ket{n_\mathrm{in},\ell_\mathrm{in},s}.
\end{align}

\noindent When that wavepacket passes through an SPP with an integer value of topological charge $q$, its OAM is increased by $q \hbar$ \cite{spinorbit}:

\begin{align}
\ket{\Psi_\mathrm{SPP}}=\hat{U}_\mathrm{SPP}\ket{\Psi_\mathrm{in}}=\sum_{n=0}^\infty C_{n,\ell_\mathrm{in}+q}\ket{n,\ell_\mathrm{in}+q,s}.
\label{Eqn:psispp}
\end{align}

\noindent The coefficients $C_{n,\ell_\mathrm{in}+q}$ are explicitly derived in  \cite{spinorbit}. Thus an SPP may be used to vary the azimuthal quantum number.

\begin{figure}
\center
\includegraphics[width=\linewidth]{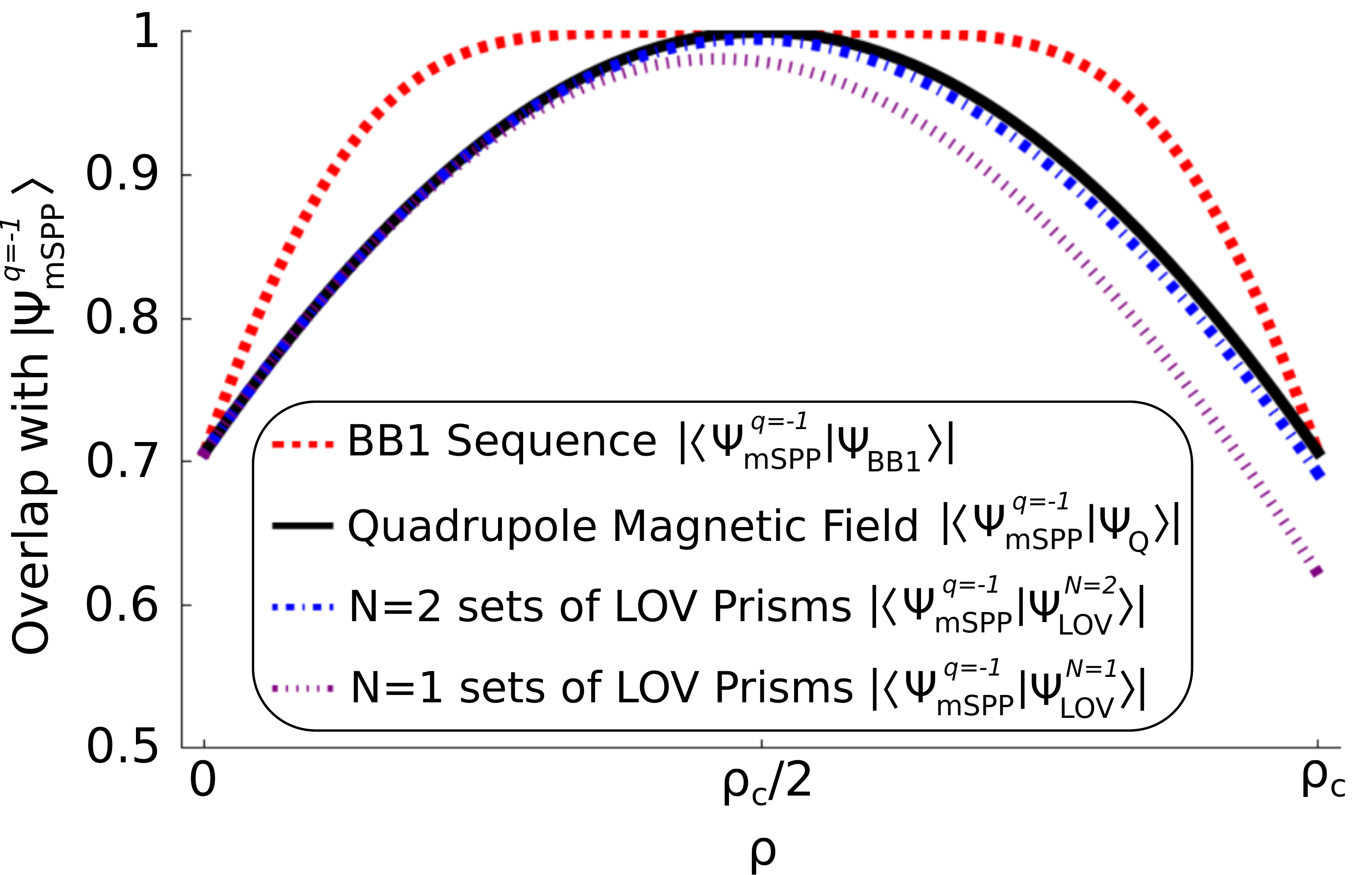}
\caption{Overlap as a function of radius between the maximally entangled spin-orbit state $\ket{\Psi^{q=-1}_{\mathrm{mSPP}}}$ and output states produced by the following methods: (red) the BB1 sequence, $\ket{\Psi_\mathrm{BB1}}$; (black) the quadrupole, $\ket{\Psi_\mathrm{Q}}$;  (blue) the N=2 sets of LOV prism pairs, $\ket{\Psi_\mathrm{LOV}^{N=2}}$; (purple) the N=1 sets of LOV prism pairs, $\ket{\Psi_\mathrm{LOV}^{N=1}}$. In each of these cases, $\rho_\mathrm{c}=1.82\sigma_\perp$. Each lattice cell of $\ket{\Psi_\mathrm{LOV}^{N=1}}$ is shown to be a good approximation of $\ket{\Psi_\mathrm{Q}}$, and the approximation is improved by reapplying the LOV operator. It is also shown that the $\ket{\Psi_\mathrm{BB1}}$ has a larger range of radii than $\ket{\Psi_\mathrm{Q}}$ for which the spin and OAM are maximally entangled.}
 \label{Fig:Fidelity}
\end{figure}

\section{\label{sec:level1}Methods of Generating spin-orbit states}
\subsection*{Method 1: Magnetic Spiral Phase Plate}

Neutrons are spin$-1/2$ particles, and therefore the spin provides a two-level DOF. A ``spin-orbit'' state is one in which spin and OAM are correlated. In this paper we specifically consider states where the two spin eigenstates are correlated with different OAM states:

\begin{align}
\ket{\Psi_\mathrm{SO}}=\frac{1}{\sqrt[]{2}}(\ket{n_\uparrow,\ell_\uparrow,\uparrow}+e^{i\beta}\ket{n_\downarrow,\ell_\downarrow,\downarrow}),
\label{Eqn:psiso}
\end{align} 

\noindent where $\ell_\uparrow\neq\ell_\downarrow$, and $\beta$ is an arbitrary phase. This state may be prepared by taking an incoming beam in a coherent superposition of spin up and spin down states (for convenience we shall choose the $\hat{z}$ axis to be the spin quantization axis and that $n_\mathrm{in}=\ell_\mathrm{in}=0$):

\begin{align}
\ket{\Psi_\mathrm{in}}=\frac{1}{\sqrt[]{2}}\ket{0,0}(\ket{\uparrow_z}+\ket{\downarrow_z}),
\end{align}
and passing it through an SPP made out of a magnetic material. When such an SPP is magnetized along the spin quantization axis, its operator can be expressed as
\begin{align}
\hat{U}_\mathrm{mSPP}=e^{i[Nb_\mathrm{c}\lambda h(\phi)+ Nb_\mathrm{m}\lambda h(\phi)\hat{\sigma}_z]}.
\label{Eqn:psimspp}
\end{align} 
\noindent where $b_\mathrm{m}$ is the neutron magnetic scattering length of the material \cite{Sears}, and $\hat{\sigma}_z$ is the Pauli spin operator. Consider an SPP which is fabricated from a material whose nuclear and magnetic scattering lengths are equal, $b_\mathrm{c}=b_\mathrm{m}$. Then the phase acquired by one spin state would be $\alpha_\uparrow(\phi)=-N(b_\mathrm{c}-b_\mathrm{m})\lambda h(\phi)=0$ and that of the other $\alpha_\downarrow(\phi)=-N(b_\mathrm{c}+b_\mathrm{m})\lambda h(\phi)=\beta+q\phi$, where now $q=-Nb_\mathrm{c}\lambda h_\mathrm{s}/\pi$ and $\beta=-2Nb_\mathrm{c}\lambda h_\mathrm{0}$. Using this magnetic SPP, spin-orbit states may be generated in the form of:
\begin{align}\nonumber
\ket{\Psi^q_\mathrm{mSPP}}&=\hat{U}_\mathrm{mSPP}\ket{\Psi_\mathrm{in}}\\
&=\frac{1}{\sqrt[]{2}}(\ket{0,0,\uparrow_z}+e^{i\beta}\sum_{n=0}^\infty C_{n,q}\ket{n,q,\downarrow_z}).
\label{Eqn:psimspp}
\end{align} 
The action of a $q=-1$ magnetic SPP is shown in Fig.~\ref{Fig:methods}a. For a convenient comparison with other methods of producing spin-orbit states we will set $\beta=\pi/2$ in Eq.~\ref{Eqn:psimspp}. $\ket{\Psi^q_\mathrm{mSPP}}$ possesses maximal single particle entanglement between the spin DOF and the OAM DOF as there is an equal superposition of $\ket{\ell_\uparrow,\uparrow_z}$ and $\ket{\ell_\downarrow,\downarrow_z}$ \cite{spinorbit}.

\subsection*{Method 2: Quadrupole Magnetic Field}\label{section:quad}

Spin-orbit states can also be prepared with a quadrupole magnetic field, as described in Ref.~\cite{spinorbit}. In this case the OAM is induced via a Pancharatnam-Berry geometrical phase \cite{pancharatnam1956generalized,berry1987adiabatic}. The spin-orbit state is achieved by propagating a neutron wavepacket that is  spin polarized along the $\hat{z}$-direction,
\begin{align}
\ket{\Psi_\mathrm{in}}=\ket{0,0,\uparrow_z},
\end{align}
through a quadrupole magnetic field $\vec{B}=K(-x\hat{x}+y\hat{y})$, where $K$ is the magnitude of the quadrupole magnetic field gradient. The Hamiltonian of a neutron inside a magnetic field can be written as $H=\vec{\hat{\sigma}}\cdot \vec{B}\gamma_\mathrm{n}\hbar/2$, where $\vec{\hat{\sigma}}$ is the vector of Pauli matrices $(\hat{\sigma}_x,\hat{\sigma}_y,\hat{\sigma}_z)$, and $\gamma_\mathrm{n}$ is the neutron gyromagnetic ratio \cite{codata}. The time that a neutron traveling along the $\hat{z}$ axis spends inside the magnetic field is $\tau=d/v_z$, where $d$ is the length of the quadrupole magnet and $v_z$ is the neutron velocity. By defining OAM raising and lowering operators $\hat{l}_\pm =e^{\pm i\phi}$ and spin operators $\hat{\sigma}_\pm =(\hat{\sigma}_x \pm i\hat{\sigma}_y)/2$, the quadrupole operator can be expressed as
\normalsize

\begin{align}
\hat{U}_\mathrm{Q}(\rho_\mathrm{c})
    	&= e^{-i\frac{\pi \rho}{2\rho_\mathrm{c}}[-\cos(\phi)\hat{\sigma}_x+\sin(\phi)\hat{\sigma}_y]} 
        \label{Eqn:UQ2} \\
 		&= \cos\left(\frac{\pi \rho}{2\rho_\mathrm{c}}\right) \mathbb{1}
        	+i\sin\left(\frac{\pi \rho}{2\rho_\mathrm{c}}\right)
            	\left(\hat{l}_+\hat{\sigma}_+ +\hat{l}_-\hat{\sigma}_-\right),\nonumber
\end{align}
	
\normalsize
\noindent where we have re-parametrized the quadrupole operator using the characteristic radial distance $\rho_\mathrm{c}$ at which the spin undergoes a $\pi$ rotation after passing through the quadrupole, 

\begin{align}
\rho_\mathrm{c}=\frac{\pi v_z}{\gamma_\mathrm{n} K d}.
\label{Eqn:rc}
\end{align}

\begin{figure}
\center
\includegraphics[width=\linewidth]{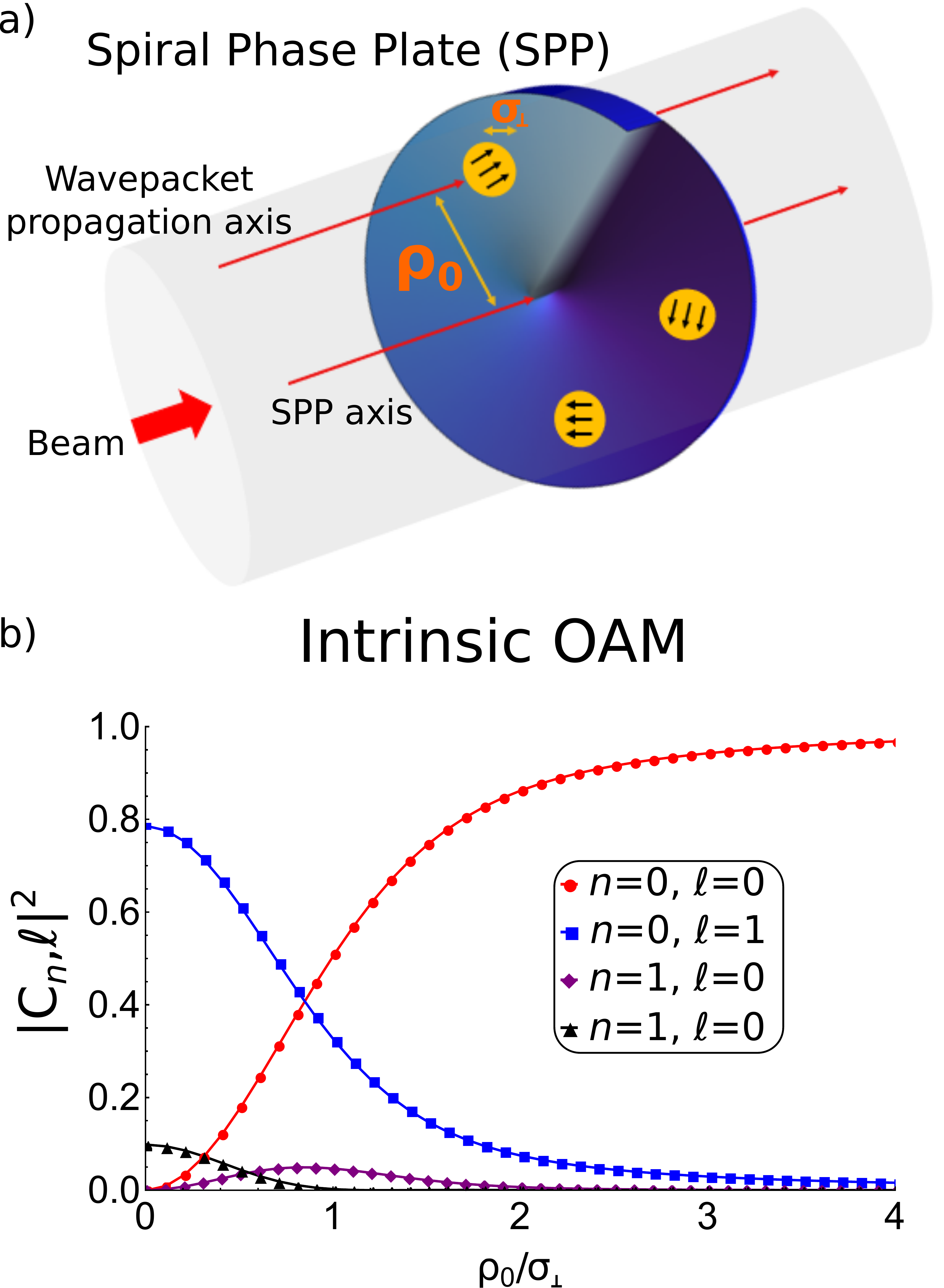}
\caption{a) As the coherence length of the neutron wavepackets is much smaller than the beam diameter, we may differentiate between ``extrinsic OAM'' calculated w.r.t. the SPP axis as the cross product of wavepacket's position and its total linear momentum, and ``intrinsic OAM'' which is associated with helical wavefronts \cite{o2002intrinsic,bliokh2015spin}. The black arrows on top of the wavepackets indicate the direction of the induced diffraction due to the SPP. b) The probabilities of the $n=0,1$ and $\ell=0,1$ states when a neutron wavepacket with no OAM $n_\mathrm{in}=\ell_\mathrm{in}=0$ passes through an SPP with $q=1$. The probabilities are calculated w.r.t. the neutron's propagation axis and they are plotted as a function of the rescaled distance from the center of the SPP, $\rho_\mathrm{0}/\sigma_\bot$, where $\rho_\mathrm{0}$ is the distance between the SPP axis and the wavepacket's propagation axis, and $\sigma_\bot$ is the transverse coherence length of the wavepacket. }
 \label{Fig:coeffs}
\end{figure}

The state after the quadrupole can be expanded in the basis functions of Eq.~\ref{basis} as
\small
\begin{align}\nonumber
\ket{\Psi_\mathrm{Q}}&=\hat{U}_\mathrm{Q}\ket{\Psi_\mathrm{in}}\\&=\frac{e^{-\xi^2/2}}{\sqrt{\pi\sigma^2_\perp}}\left[\cos\left(\frac{\pi \rho}{2\rho_\mathrm{c}}\right) \ket{\uparrow_z}+ie^{-i\phi}\sin\left(\frac{\pi \rho}{2\rho_\mathrm{c}}\right)\ket{\downarrow_z}\right]\notag\\&=
\sum_{n=0}^{\infty}( C_{n,0,\uparrow_z}\ket{n,0,\uparrow_z}+
 iC_{n,-1,\downarrow_z }\ket{n,-1,\downarrow_z }),
\label{Eqn:psiq}
\end{align}
\normalsize
where the coefficients $C_{n,\ell,s}$ are explicitly derived in Ref.~\cite{spinorbit}. There it was also shown that to maximize the single particle entanglement between the spin and OAM the quadrupole magnet should be of such strength and length as to produce a spin flip over 1.82 times the coherence length of the wavepacket, that is $\rho_\mathrm{c}=1.82\sigma_\bot$. 

The action of the quadrupole magnet is shown in Fig.~\ref{Fig:methods}b. It can be observed that the intensity profile of the spin state which is correlated to the OAM is now a ring shape.

\subsection*{Method 3: BB1 Sequence}

After a neutron wavepacket passes through a quadrupole magnetic field, the maximally entangled spin-orbit state, given by $\ket{\Psi^{q=-1}_\mathrm{mSPP}}$ (see Eq.\ref{Eqn:psimspp}), occurs for $\rho=\rho_\mathrm{c}/2$. However, the range of maximal entanglement can be increased by using a sequential chain of appropriately oriented quadrupole magnets. We will see that this results in the ability to increase the width of the ideal ring filter without significantly affecting the amount of spin-orbit entanglement, boosting post-selection performance. 
To begin, notice that the situation with a single quadrupole magnet resembles a standard over/under-rotation pulse error in spin physics \cite{levitt_composite_1986}:
with a fixed azimuthal coordinate $\phi$, as the radial coordinate deviates from the ideal value $\rho=\rho_\mathrm{c}/2$, the spin undergoes a rotation about the $\hat{\phi}$ axis with a rotation angle greater or less than $\pi/2$.
The amount of such over/under-rotation is fixed for a given value of $\rho$.

\begin{figure*}
\center
\includegraphics[width=\linewidth]{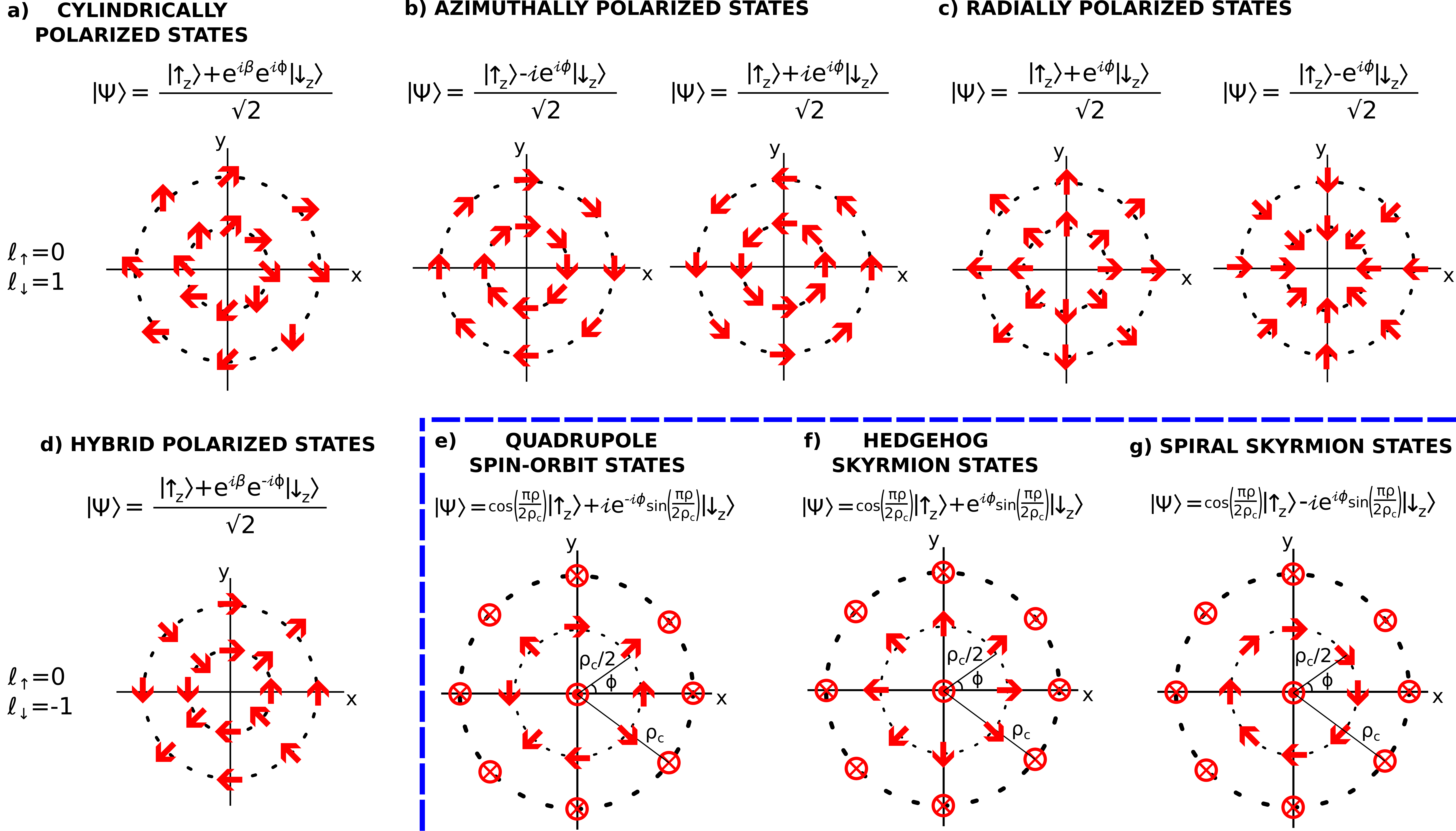}
\caption{The spin orientation (red arrows) of the spin-orbit states with a coupling between $\ell_\uparrow=0$ and $\ell_\downarrow=\pm 1$, where the $\hat{z}$ axis points out of the page. In analogy to optical OAM terminology, we may classify four categories of spin-orbit states with radially independent spin orientations: a) ``cylindrically polarized states'' where the spin orientation is given by $\vec{P}=\cos(\beta)\hat{r}+\sin(\beta)\hat{\phi}$, where $\beta$ is an arbitrary phase; b) ``azimuthally polarized states'' which are a subset of cylindrically polarized states where $\vec{P}=\pm\hat{\phi}$; c) ``radially polarized states'' which are a subset of cylindrically polarized states where $\vec{P}=\pm\hat{r}$; and d) ``hybrid polarization states'' where $\vec{P}=\sin(2\phi+\beta)\hat{r}+\cos(2\phi+\beta)\hat{\phi}$, where $\beta$ is an arbitrary phase. Note that all of the states with a certain $\{\ell_\uparrow,\ell_\downarrow\}$ differ by a phase on the spin DOF. The preparation techniques shown in Fig.~\ref{Fig:methods} can also produce spin-orbit states with radially dependent spin orientations. The main three categories are: e) quadrupole spin-orbit states as described by Eq.~\ref{Eqn:psiq}; f) hedgehog skyrmion states; and g) spiral skyrmion states. An array of any of these three states can be obtained via the appropriate LOV prism pair combination. 
}
 \label{Fig:vectorbeams}
\end{figure*}

To increase robustness to these errors we consider the Broad-Band$_1$ (BB1) composite pulse ~\cite{wimperis_broadband_1994} which can be implemented by sequential quadrupoles with different strengths and orientations. This particular composite sequence is considered because of its robust performance while using only four quadrupole magnets. It is important to note that applying the quadrupole operator repeatedly $N$ times does not take the orbital quantum numbers outside the $\ell=0,\pm 1$ values. That is, $[\hat{U}_\mathrm{Q}(\rho_\mathrm{c})]^N \ket{\Psi_\mathrm{in}}=\hat{U}_\mathrm{Q}(\rho_\mathrm{c}/N) \ket{\Psi_\mathrm{in}}$, where the quadrupole operator $\hat{U}_\mathrm{Q}(\rho_\mathrm{c})$ was defined in Eq.~\ref{Eqn:UQ2}.
However, the standard magnetic quadrupole can be  rotated by an angle $\delta$ about the $\hat{z}$ axis. In this case its interaction is described by the modified operator, $\hat{\mathcal{U}}_\mathrm{Q}(\rho_\mathrm{c},\delta)=e^{-i \frac{\delta}{2}\hat{\sigma}_z}\hat{U}_\mathrm{Q}(\rho_\mathrm{c})e^{i \frac{\delta}{2}\hat{\sigma}_z}$, and the BB1 sequence results in the output state

\begin{align}
\ket{\Psi_\mathrm{BB1}}=\hat{\mathcal{U}}_\mathrm{Q}\left(\frac{\rho_\mathrm{c}}{2},\delta_1\right)\hat{\mathcal{U}}_\mathrm{Q}\left(\frac{\rho_\mathrm{c}}{4},\delta_2\right)
  \notag\\
\hat{\mathcal{U}}_\mathrm{Q}\left(\frac{\rho_\mathrm{c}}{2},\delta_1\right) \hat{\mathcal{U}}_\mathrm{Q}\left(\rho_\mathrm{c},0\right)\ket{\Psi_\mathrm{in}},
\end{align}

\noindent where $\delta_1=\cos^{-1}(-1/8)$ and $\delta_2=3\delta_1$.
These angles were tuned to eliminate 1\textsuperscript{st} and 2\textsuperscript{nd} order over/under-rotation errors~\cite{wimperis_broadband_1994}. 

To quantitatively compare $\ket{\Psi_\mathrm{BB1}}$ with $\ket{\Psi_\mathrm{Q}}$ we can look at their overlap with the maximally entangled spin orbit state $\ket{\Psi^{q=-1}_\mathrm{mSPP}}$ of Eq.~\ref{Eqn:psimspp}. The overlap between two states $\ket{\Psi_1}$ and $\ket{\Psi_2}$ is given by $|\braket{\Psi_1|\Psi_2}|$, and it is a measure of the closeness of two quantum states, with a value of unity for identical states.  Fig.~\ref{Fig:Fidelity} shows $|\braket{\Psi^{q=-1}_\mathrm{mSPP}|\Psi_\mathrm{BB1}}|$ and $|\braket{\Psi^{q=-1}_\mathrm{mSPP}|\Psi_\mathrm{Q}}|$ as a function of radius. It is clear that $\ket{\Psi_\mathrm{BB1}}$ has a larger range of radii for which the spin and OAM are maximally entangled. This can also be observed in the intensity profile of $\braket{\downarrow_z|\Psi_\mathrm{BB1}}$ that is plotted in Fig.~\ref{Fig:methods}c, where the inner dark region is smaller than that of Fig.~\ref{Fig:methods}b.

\subsection*{Spin-Orbit states with higher order OAM}

The quadrupole magnetic field method described above takes a spin-polarized input state with $\ell_{\uparrow}=\ell_{\downarrow}=0$ and outputs a spin-orbit state with $\ell_{\uparrow}=0$ and $\ell_{\downarrow}=\pm 1$. We now consider situations where the spin-orbit correlations involve higher order OAM values. With spin-orbit states generated via the magnetic SPP, this is a trivial matter of using a $|q|>1$. For quadrupole magnetic fields the following sequence of $j$ pulses may be used:

\small
\begin{align}\label{Eqn:higherq}
\ket{\Psi^j_\mathrm{Q}}=\left(\hat{U}_\mathrm{Q}(\rho_\mathrm{c})e^{-i\frac{\pi}{2}\hat{\sigma}_x}\ket{\downarrow_z}\bra{\downarrow_z}\right)^j\hat{U}_\mathrm{Q}(\rho_\mathrm{c})\ket{0,0,\uparrow_z}\\\nonumber
=\frac{e^{-\xi^2/2}}{\sqrt{\pi\sigma^2_\perp}}[\cos\left(\frac{\pi \rho}{2 \rho_\mathrm{c}}\right)\sin^j\left(\frac{\pi \rho}{2 \rho_\mathrm{c}}\right)\ket{-j,\uparrow_z}\\\nonumber+i\sin^{j+1}\left(\frac{\pi \rho}{2 \rho_\mathrm{c}}\right)\ket{-(j+1),\downarrow_z}]
\end{align}
\normalsize

\noindent where $\ket{\downarrow_z}\bra{\downarrow_z}$ is the projection operator for a spin-down state. The $j=0$ case corresponds to the spin-orbit state produced via a quadrupole magnetic field as described in Eq.\ref{Eqn:psiq}. For $j>1$, both $\ket{\uparrow_z}$ and $\ket{\downarrow_z}$ are correlated to higher order OAM values, and the intensity profiles of $\braket{\uparrow_z|\Psi^j_\mathrm{Q}}$ and $\braket{\downarrow_z|\Psi^j_\mathrm{Q}}$ are both ring shapes.

\section{Intrinsic and Extrinsic OAM}

\begin{figure*}
\center
\includegraphics[width=\linewidth]{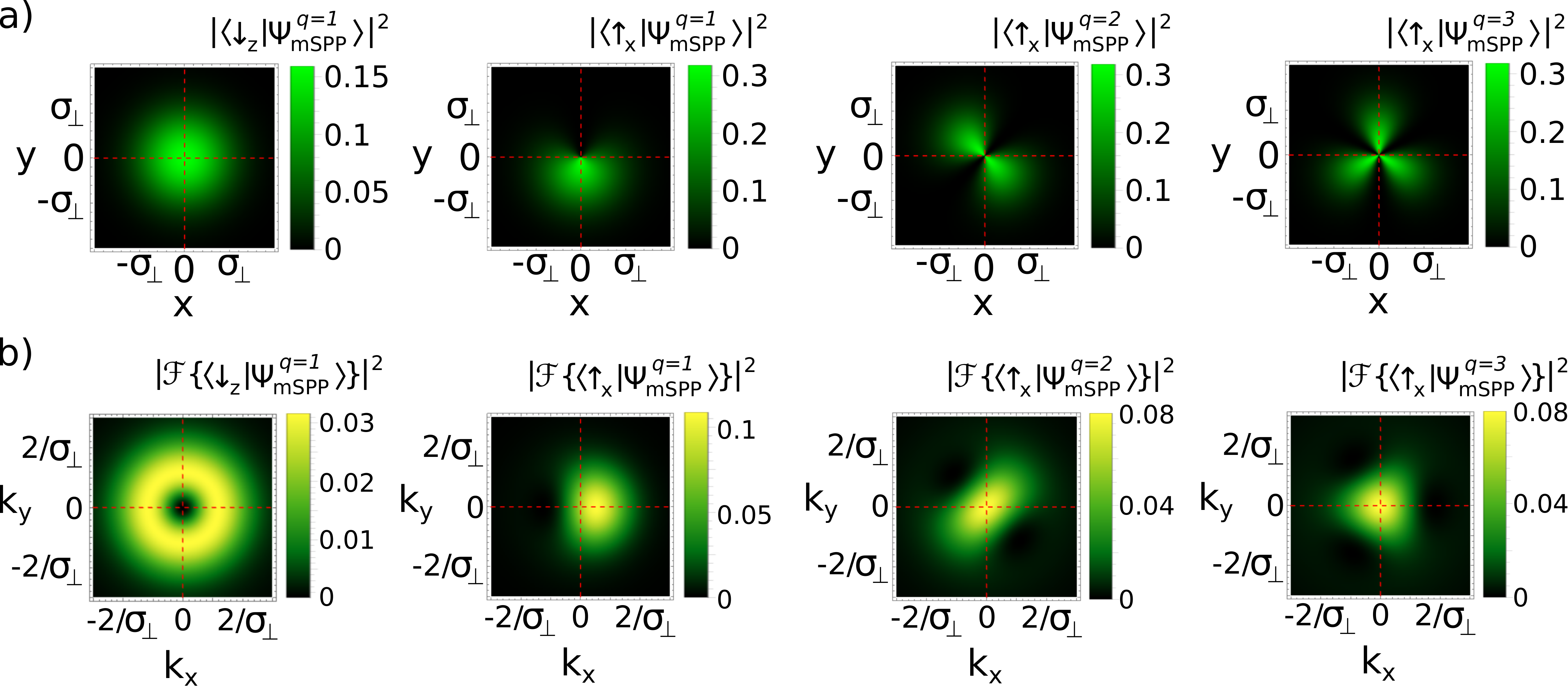}
\caption{The two parameters of spin-orbit states, $\beta$ and $\Delta\ell=\ell_\uparrow-\ell_\downarrow$ (see Eq.~\ref{Eqn:psiso}), can be characterized by post-selecting on a perpendicular spin direction and obtaining: a) the 2D intensity profile or b) the 2D momentum distribution. The first two columns are for the state after a magnetic SPP with q=1, the third column is for the state after a magnetic SPP with q=2, and the last column is for the state after a magnetic SPP with q=3. The order of rotational symmetry of the 2D intensity and momentum profiles is equal to $|\Delta\ell|=|q|$ (as we set $\ell_\downarrow=0$ for convenience). Applying a spin rotation along $\hat{\sigma}_z$ before the spin mixing effectively rotates the resulting 2D profiles. The direction of rotation determines the sign of q, and the initial azimuthal offset determines $\beta$ at the detector.}
 \label{Fig:measuringOAM}
\end{figure*}

Heretofore, we have discussed neutron wavepackets for which the propagation axis coincides with the SPP or quadrupole axis. In this case, the SPP/quadrupole axis defines the OAM quantization axis. However, neutron beams are typically an incoherent superposition of neutron wavepackets, where the neutron beam diameter is  between $10^{-1}$ m and $10^{-4}$ m, and the transverse coherence length of the neutron wavepackets, $\sigma_\perp$, is of the order of $10^{-5}$ m to $10^{-9}$ m \cite{petrascheck1988coherence,pushin2008measurements,sarenac2017three}. 

In studies of optical OAM a distinction is made between ``extrinsic OAM'' and ``intrinsic OAM'' \cite{o2002intrinsic,bliokh2015spin}. One can extend this distinction to the case of neutron beams. Extrinsic OAM is the orbital angular momentum centered about the SPP/quadrupole axis and it is given by the cross product of wavepacket's position and its total linear momentum; intrinsic OAM, usually associated with helical wavefronts, is the orbital angular momentum represented by $\ell$. The intrinsic OAM does not depend upon the position of the axis, provided  that the axis is parallel to the propagation axis \cite{berry1998paraxial}. This is depicted on Fig.~\ref{Fig:coeffs}a which shows that a helical wavefront is induced only for the wavepacket whose propagation axis coincides with the SPP axis. 

Consider a neutron wavepacket with $n_\mathrm{in}=\ell_\mathrm{in}=0$ and which is centered on $(\rho_\mathrm{0},\phi_\mathrm{0})$: 

\begin{align}
\ket{\Psi_\mathrm{o}}=\frac{1}{\sqrt[]{\pi\sigma_\perp^2}}e^{-\frac{\rho^2+\rho_\mathrm{0}^2-2\rho\rho_\mathrm{0}\cos(\phi-\phi_\mathrm{0})}{2\sigma_\perp^2}},
\end{align}

\noindent After passing through an SPP which is centered at $\rho=0$, the expectation value of OAM about the SPP axis is:

\begin{align}\nonumber
\textless\hat{L}_z\textgreater=\int_0^\infty d\rho \int_0^{2\pi}d\phi\: \rho\bra{\Psi_\mathrm{o}}\hat{U}_\mathrm{SPP}^\dag\\\left(-i\hbar\frac{\partial}{\partial\phi}\right)\hat{U}_\mathrm{SPP}\ket{\Psi_\mathrm{o}}=\hbar q.
\end{align}

\noindent Therefore all wavepackets in the output beam acquire a well defined mean OAM relative to the SPP axis. Such wavepackets are diffracted in the transverse direction, such that the induced external OAM relative to the SPP axis is independent of their location: 
\begin{align}
L_z=\vec{r}\times\vec{p}=\rho_\mathrm{0}\hbar k_\perp=\hbar q,
\end{align}
\noindent where  $k_\perp=q/\rho_\mathrm{0}$ is induced by the SPP (in Fig.~\ref{Fig:coeffs}a the diffraction direction is depicted with black arrows). 

On the other hand, as shown in Fig.~\ref{Fig:coeffs}b, the intrinsic OAM of a neutron wavepacket quickly vanishes as the the wavepacket's propagation axis is displaced from the center of the SPP. The intrinsic OAM of the output beam has a Gaussian dependence to the displacement from the center of the SPP \cite{oemrawsingh2004intrinsic}.

\section{Lattices of Spin-Orbit states}

For material studies there is a need for methods to generate lattices of neutron spin-orbit states where the lattice constants are matched to the characteristic length scales of topological and chiral materials.  We show how this may be achieved via a sequence of magnetic field gradients. 

A lattice of optical spin-orbit states can be produced using sets of specially arranged birefringent prism pairs denoted as ``LOV prism pairs''. This procedure was demonstrated for the polarization DOF of electromagnetic waves in Ref.~\cite{sarenac2017generation}. Here, we consider the spin DOF of matter-waves. 

\begin{figure*}
\center
\includegraphics[width=\linewidth]{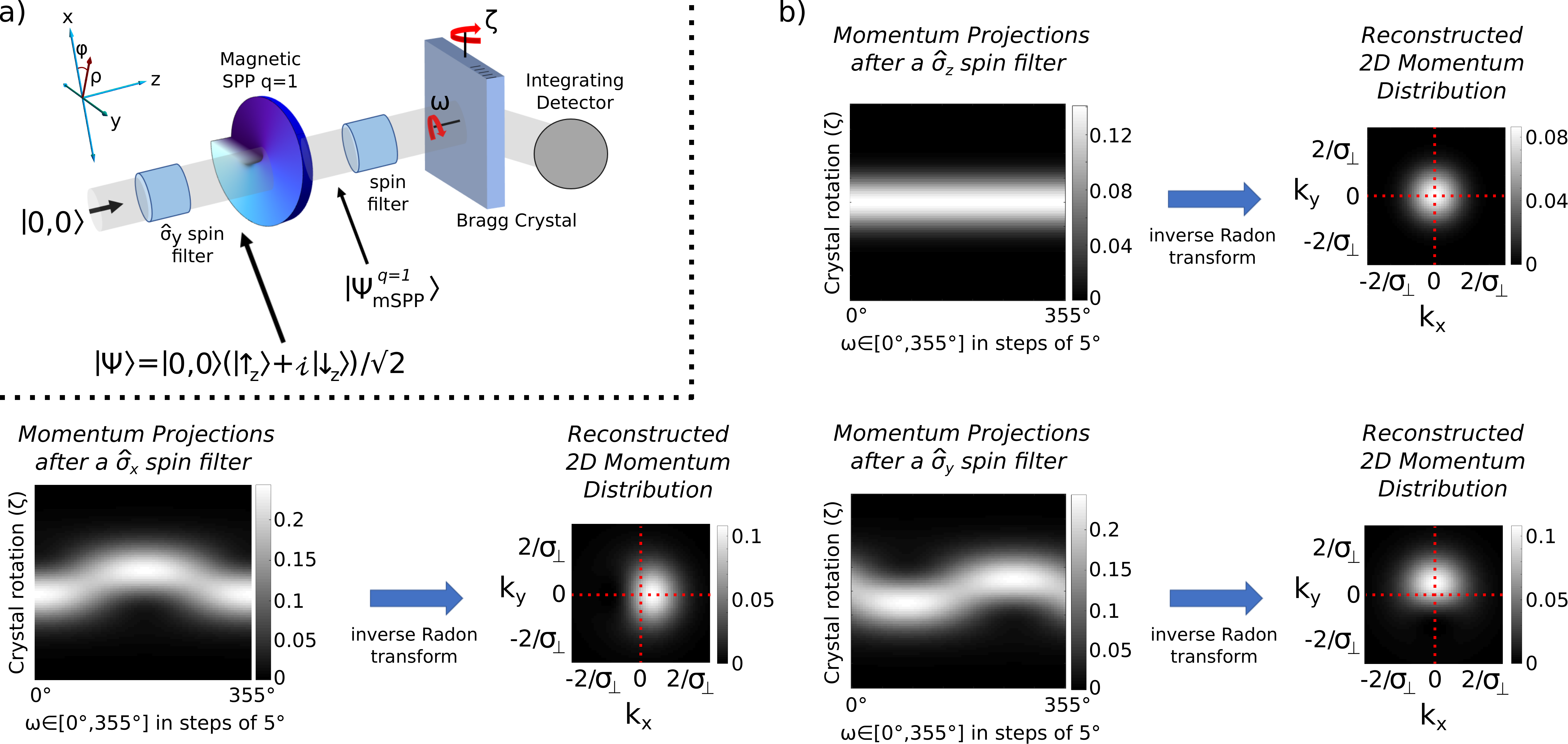}
\caption{When post-selecting onto a perpendicular spin eigenbasis of a spin-orbit state the OAM manifests itself as an asymmetry in the 2D momentum distribution (see Fig.~\ref{Fig:measuringOAM}). a) Proposed experiment to map out the 2D momentum distribution of a neutron spin-orbit state by measuring the momentum projections via a Bragg crystal. This allows for analysis of the beam's OAM components by mapping out the momentum distribution. b) We may assemble the momentum projections at each $\omega$ obtained by rotating the Bragg crystal around the crystal plane direction. The 2D momentum distribution is obtained from the projection curves via the inverse Radon transform. In the examples depicted we perform the inverse Radon transform on 36 equally spaced slices of $\omega\in[0\degree,175\degree]$ and reconstruct the 2D momentum distribution. 
}
 \label{Fig:radon}
\end{figure*}

The method to produce lattices of spin-orbit states is motivated by applying the Suzuki-Trotter expansion to Eq.~\ref{Eqn:UQ2}: 

\begin{align}
	e^{i\frac{\pi}{2\rho_\mathrm{c}}(x\hat{\sigma}_x-y\hat{\sigma}_y)}
    	= \lim_{N\to\infty} 
        	(e^{i\frac{\pi}{2\rho_\mathrm{c}N}x\hat{\sigma}_x}
            e^{-i\frac{\pi}{2\rho_\mathrm{c}N}y\hat{\sigma}_y})^N.
	\label{Eqn:Lie}
\end{align}
We can see that $N$ set of perpendicular linear magnetic gradients approximates the quadrupole operator. Choosing that the operators be independent of $N$, we define the linear magnetic gradient operator as

\begin{align}
	\hat{U}_{\phi_\mathrm{g},\phi_\mathrm{m}}=e^{-i\frac{\pi }{2 \rho_\mathrm{c}}[x\cos\phi_\mathrm{g}+y\sin\phi_\mathrm{g}][\hat{\sigma}_x\cos\phi_\mathrm{m}+\hat{\sigma}_y\sin\phi_\mathrm{m}]} 
	\label{gradientoperators}
\end{align}

\noindent where $\phi_\mathrm{g}$ ($\phi_\mathrm{m}$) indicates the gradient (magnetic field) direction in the $x-y$ plane. For spin$-1/2$ particles one way to approximate the magnetic linear gradient operators is with magnetic prisms as shown in Fig.~\ref{Fig:methods}d. These are matter-wave analogous of the LOV prism pairs introduced in Ref.~\cite{sarenac2017generation}. The general LOV operator can be expressed as:

\begin{align}
	\hat{U}_\mathrm{LOV}^N
    	=(\hat{U}_{\phi_\mathrm{g},\phi_\mathrm{m}} \hat{U}_{\phi_\mathrm{g}\pm\frac{\pi}{2},\phi_\mathrm{m}\pm\frac{\pi}{2}})^N,
	\label{Eqn:gradsoperator}
\end{align}

\noindent and the corresponding beams with lattices of spin-orbit states are given by:

\begin{align}
	\ket{\Psi_\mathrm{LOV}^N}
    	=(\hat{U}_{\phi_\mathrm{g},\phi_\mathrm{m}} \hat{U}_{\phi_\mathrm{g}\pm\frac{\pi}{2},\phi_\mathrm{m}\pm\frac{\pi}{2}})^N\ket{\Psi_\mathrm{in}}.
	\label{Eqn:psigrads}
\end{align}

This process is shown in Fig.~\ref{Fig:methods}d for $(\hat{U}_{\pi,0} \hat{U}_{\frac{\pi}{2},\frac{\pi}{2}})^2$ and $\ket{\Psi_\mathrm{in}}=\ket{\uparrow_z}$, where the output beam is a lattice of spin-orbit states with $\ell_\uparrow=0$ and $\ell_\downarrow=-1$. 
The orientations of the gradient operators give us the possibility of producing lattices of spin-orbit states with positive and negative values of OAM. For example, $(\hat{U}_{0,0} \hat{U}_{\frac{\pi}{2},\frac{\pi}{2}})^2$ applied to an incoming state of $\ket{\Psi_\mathrm{in}}=\ket{\uparrow_z}$ produces an output beam with a lattice of spin orbit states with $\ell_\uparrow=0$ and $\ell_\downarrow=1$. Note that this particular gradient sequence approximates the action of a monopole magnetic field geometry. Furthermore, we can obtain lattices of spin-orbit states with higher order OAM values by substituting the LOV operator, $\hat{U}_\mathrm{LOV}^N$, in place of the quadrupole operators, $\hat{U}_\mathrm{Q}(\rho_\mathrm{c})$, in Eq.~\ref{Eqn:higherq}. 

Due to the periodic nature of the linear gradient operators, the spin-orbit states in these beams form a two-dimensional array with a lattice constant of

\begin{align}
	a=\frac{2\pi v_z}{ \gamma_\mathrm{n} |B|  \tan(\theta)}
	\label{Eqn:const}
\end{align}

\noindent where $|B|$ is the magnitude of the magnetic field and  $\theta$ is the inclination angle of the LOV prism pairs. In Fig.~\ref{Fig:methods}d the phase and intensity profiles of the polarization state which is correlated with the OAM illustrate the lattice structure. The number of well defined intensity rings in a lattice cell is equal to $N/2$, where $N$ is the number of LOV prism pairs. Therefore, $N$ provides control over the mean radial quantum number $n$ in the lattice cells \cite{sarenac2017generation}. 

\section{Polarization Geometries of Spin-Orbit States}

Following the nomenclature of polarization correlated OAM states \cite{rubinsztein2016roadmap}, we classify neutron spin-orbit states according to their spin orientation profile. There are four categories of spin-orbit states with radially independent spin orientations as shown in Fig.~\ref{Fig:vectorbeams}a-d. They are: 
\begin{enumerate}[label=(\alph*)]
\item ``cylindrically polarized states'' where the spin orientation is given by $\vec{P}=\cos(\beta)\hat{\rho}+\sin(\beta)\hat{\phi}$, where $\beta$ is an arbitrary phase; 
\item``azimuthally polarized states'' which are a subset of cylindrically polarized states where $\vec{P}=\pm\hat{\phi}$; 
\item``radially polarized states'' which are a subset of cylindrically polarized states where $\vec{P}=\pm\hat{r}$; and 
\item``hybrid polarization states'' where $\vec{P}=\sin(2\phi+\beta)\hat{r}+\cos(2\phi+\beta)\hat{\phi}$, where $\beta$ is an arbitrary phase. 
\end{enumerate}

The simplest method to generate any of those four states is to pass an appropriate input state into the magnetic SPP of $q=\pm1$, as the four categories arise when $\Delta\ell=\ell_\uparrow-\ell_\downarrow=\pm 1$. The optical spin-orbit states with analogous polarization orientation geometries are not characterized by $\Delta\ell=\pm 1$. This difference comes from the fact that on the Poincar\'e\ sphere that describes optical polarization, any two antipodal points refer to orthogonal polarization directions; while on the Bloch sphere that describes the spin-$1/2$ state, any two antipodal points refer to anti-parallel spin directions. 

We consider a spin-orbit state for which one orbital quantum number is zero and the other $\pm 1$. When  $\ell_\uparrow=0$ the hybrid polarized states of Fig.~\ref{Fig:vectorbeams}d possess $\{\ell_\uparrow=0,\ell_\downarrow=-1\}$, and the cylindrically polarized states possess $\{\ell_\uparrow=0,\ell_\downarrow=1\}$. All of the states with given $\{\ell_\uparrow,\ell_\downarrow\}$ differ by a phase on the spin DOF. This phase can be directly varied by an external magnetic field along the spin quantization axis, $B_z$. For $\ell_\downarrow=0$ the hybrid polarized states possess $\{\ell_\uparrow=1,\ell_\downarrow=0\}$ while the cylindrically polarized states possess $\{\ell_\uparrow=-1,\ell_\downarrow=0\}$. Hence a $\pi$ spin rotation around $\hat{\sigma}_{\perp}$ can be used to transform a state with hybrid polarization geometry into a state with cylindrical polarization geometry (and vice versa), but not to change $\Delta\ell$.

The preparation techniques shown in Fig.~\ref{Fig:methods} can also produce spin-orbit states with radially dependent spin orientations. The main three categories are shown in Fig.~\ref{Fig:vectorbeams}: e) quadrupole spin-orbit state; and two skyrmion-like states: f) hedgehog and g) spiral. The described rules for radially independent spin-orbit states also apply to these radially dependent spin-orbit states. The quadrupole spin-orbit state is described by Eq.~\ref{Eqn:psiq}, while a lattice of any of these three categories of states can be obtained via an appropriate LOV prism pair combination. 

\section{Characterization of spin-orbit states}

Generally speaking, determining a neutron beam's OAM is relatively difficult due to the low flux and small spatial coherence length. One possible method is to prepare the OAM beam in one arm of an interferometer, which will yield an output beam that is a coherent superposition of the OAM beam and a reference beam carrying no OAM \cite{Dima2015}. The 2D intensity profile of the output beam will possess a helical structure whose order of rotational symmetry quantifies the induced OAM. In principle, it would also be possible to verify the OAM of a neutron beam by transferring the OAM from the beam to an absorbing object or particle, which would then rotate around the OAM axis as a result. This would be analogous to the optical experiments \cite{Andersen2006,he1995direct,friese1996optical}, though the available low neutron fluxes make this experiment unpractical. 

The spin-orbit states described by Eq.~\ref{Eqn:psiso} are characterized by two parameters of interest: $\Delta\ell=\ell_\uparrow-\ell_\downarrow$ and the phase factor $\beta$. Here we describe two robust and relatively simple methods to determine those parameters. However, it is important to keep in mind that $\beta$ will be varied by the background quantization magnetic field $B_z$.

\subsection{Mapping the 2D intensity profile after spin mixing}\label{subsection:kdist}

The two paths of a Mach-Zehnder interferometer are isomorphic to a two-level quantum system such as the spin$-1/2$ DOF. Therefore after a mixing in the spin DOF, the spin dependent 2D intensity profiles will possess a helical structure which quantifies the induced OAM. For simplicity consider the spin-orbit state $\ket{\Psi^q_\mathrm{mSPP}}$ (Eq.~\ref{Eqn:psispp}). The two-dimensional intensity, post-selected on a particular spin direction $\ket{s}$, is given by

\begin{align}
I(x,y)=|\braket{s|\Psi^q_\mathrm{mSPP}}|^2
\label{Eqn:psisoMeasure}
\end{align} 

\noindent Without spin mixing, i.e. post-selecting on $\ket{\uparrow_z}$ or $\ket{\downarrow_z}$, the resulting 2D intensity profile is a Gaussian in both cases, which does not reveal any OAM structure.

To determine the induced OAM on the $\ket{\downarrow_z}$ component we would need to post-select on a perpendicular spin direction. The 2D intensity profiles projected onto $\ket{\uparrow_x}$, given by $|\braket{\uparrow_x|\Psi^q_\mathrm{mSPP}}|^2$,  are shown in Fig.~\ref{Fig:measuringOAM}a) for magnetic SPPs with $q=1,2,3$. These are identical to the expected profiles obtained via the interferometric measurement described above.

The order of rotational symmetry of the 2D intensity profiles is equal to $|\Delta\ell|=|\ell_\uparrow-\ell_\downarrow|=|q|$. Applying a spin rotation along $\hat{\sigma}_z$ before the spin mixing effectively rotates the resulting 2D intensity profile. The direction of rotation determines the sign of $q$. The initial azimuthal offset determines $\beta$ at the detector.

\subsection{Mapping the 2D momentum distribution after spin mixing}

Another method to characterize spin-orbit states is to measure their 2D momentum distribution. The 2D momentum distribution, post-selected on a particular spin direction $\ket{s}$, is given by
\begin{align}
P(k_x,k_y)=|\mathcal{F}\{\braket{s|\Psi^q_\mathrm{mSPP}}\}|^2
\label{Eqn:psisoMeasure}
\end{align} 

\noindent where $\mathcal{F}\{\}$ is the Fourier transform.
If we apply spin filters along the spin eigenbasis of $\ket{\Psi^q_\mathrm{mSPP}}$, i.e. along $\ket{\uparrow_z}$ or $\ket{\downarrow_z}$, then the 2D momentum distribution of $\braket{\uparrow_z|\Psi^q_\mathrm{mSPP}}$ would be a Gaussian profile indicative of the prepared incoming state carrying no OAM, and that of $\braket{\downarrow_z|\Psi^q_\mathrm{mSPP}}$ would be a ring shape. However, the ring-shaped momentum distribution does not uniquely define an OAM beam; for example, it is possible to have a radially diverging beam which has a ring-shaped 2D momentum distribution.

If we post-select on a perpendicular spin axis then the spin-orbit coupling breaks the symmetry of the 2D momentum distribution profile as shown in Fig.~\ref{Fig:measuringOAM}b). Therefore we propose a method to characterize the spin orbit states by mapping out their 2D momentum distribution after spin filtering along a perpendicular spin axis. 

In this method as well, the order of rotational symmetry of the 2D momentum profiles is equal to $|\Delta\ell|=|\ell_\uparrow-\ell_\downarrow|=|q|$. Applying a spin rotation along $\hat{\sigma}_z$ before the spin mixing effectively rotates the resulting 2D momentum profile. The direction of rotation determines the sign of q. The initial azimuthal offset determines $\beta$ at the detector.

Allowing a state to propagate into the far field, where the intensity profile is indicative of the momentum distribution profile, is not practical with the small neutron diffraction angles induced by the OAM. A more practical method is to use a diffracting crystal and obtain momentum projection curves which can then be used to reconstruct the 2D momentum distribution. A proposed experiment is shown in Fig.~\ref{Fig:radon}a. A spin-orbit state is prepared by passing a coherent superposition of the two spin eigenstates through a magnetic SPP. The spin is then projected onto a perpendicular spin direction using a spin filter. A rotatable Bragg crystal enables a measurement of the momentum projected to the crystal plane direction. The two rotation angles $\omega$ and $\zeta$ effectively allow us to obtain the projections of the 2D momentum distribution along an arbitrary angle in the transverse plane, as shown in Fig.~\ref{Fig:radon}b).  A standard problem of medical imaging, obtaining the ``backprojection image'' (2D momentum distribution) via the ``sinogram'' (projection curves) is achieved with the inverse Radon transform \cite{smith2010introduction}. Fig.~\ref{Fig:radon}b) shows the reconstructed image obtained via 36 equally spaced projections. Note that because of the azimuthal symmetry of the spin-orbit state, rotating the spin filter of Fig.~\ref{Fig:radon}a by an angle $\omega$ and fixing the Bragg crystal orientation produces the same outcome as shown in Fig.~\ref{Fig:radon}b).

These procedures work similarly if the spin-orbit state is created via any method depicted in Fig.\ref{Fig:methods}. Note that other than the magnetic SPP, the other methods produce radial diffraction in addition to the azimuthal diffraction. However this does not change the described azimuthal asymmetry used to characterize the spin-orbit states. In fact, the asymmetry becomes even more pronounced. Therefore we proposed that an initial experiment be done with LOV prism pairs to maximize the use of the incoming beam flux and circumvent problems with small coherence lengths.

\section{Conclusion}

We have introduced and quantified new methods of preparing neutron spin-orbit states. This is a step towards general programming of the spin and quantum phase of neutron wavefronts, which addresses the fundamental limitations of neutron scattering and imaging techniques. For example, recent interest in complex topological and quantum materials \cite{nagaosa2013topological, schulz2012emergent} suggests a need for a tool with unique penetrating abilities and magnetic sensitivity. Analysis of material properties could be performed using a neutron spin-orbit lattice where the lattice constants are matched to the characteristic length scales of materials. The methods described here allow for the direct control of spin-orbit state parameters within a neutron beam. We have also proposed a method to characterize neutron spin-orbit states which overcomes the main challenges associated with low neutron flux and the neutron's small spatial coherence length.

\section{Acknowledgements}

This work was supported by the Canadian Excellence Research Chairs (CERC) program, the Natural Sciences and Engineering Research Council of Canada (NSERC) Discovery program, Collaborative Research and Training Experience (CREATE) program, the Canada  First  Research  Excellence  Fund  (CFREF), and the National Institute of Standards and Technology (NIST) Quantum Information Program.

\bibliography{OAMPRX}

\end{document}